\documentclass[twocolumn,showpacs,preprintnumbers,amsmath,amssymb]{revtex4}

\usepackage{graphicx}
\usepackage{dcolumn}
\usepackage{bm}

\begin{document}

\title{Spin Susceptibility and Superexchange Interaction in the Antiferromagnet CuO}

\author{T. Shimizu}
\email{shimizu.tadashi@nims.go.jp}
\author{T. Matsumoto}
\author{A. Goto}
\author{T. V. Chandrasekhar Rao}
\affiliation{National Institute for Materials Science, 3-13 Sakura, Tsukuba, Ibaraki 305-0003, Japan}

\author{K. Yoshimura}
\author{K. Kosuge}
\affiliation{Department of Chemistry, Kyoto University, Kyoto 606-8502, Japan}

\date{\today}

\begin{abstract}
Evidence for the quasi one-dimensional (1D) antiferromagnetism of CuO is presented in a framework of Heisenberg model.  
We have obtained an experimental absolute value of the paramagnetic spin susceptibility of CuO  by subtracting the orbital susceptibility separately from the total susceptibility through the $^{63}$Cu NMR shift measurement, and compared directly with the theoretical predictions.  
The result is best described by a 1D $S=1/2$ antiferromagnetic Heisenberg (AFH) model, supporting the speculation invoked by earlier authors.  
We also present a semi-quantitative reason why CuO, seemingly of 3D structure, is unexpectedly a quasi 1D antiferromagnet.
\end{abstract}


\maketitle
\section{Introduction}
CuO is one of the materials closely related to the high-$T_c$ cuprate superconductors, especially in light of the strong antiferromagnetic correlation in the Cu-O-Cu bonds. 
It may be surprising that unusual magnetic properties are found in CuO which consists only of the essence of cuprates, copper and oxygen ions, and has a nearly 3D structure from the viewpoint of chemical bonding.  
A close inspection of the magnetic properties in CuO may give us an insight into the underlying physics in the magnetism and superconductivity of the cuprate family.   

Extensive studies including magnetic susceptibility, specific heat, photoemission, neutron scattering and NMR measurements on CuO have revealed that (1) successive magnetic transitions are occurred at $T_{N1}=212$ K and $T_{N2}=231$ K  \cite{FBW88, YTTS89, LMJ+89}, (2) a N\'{e}el state with the easy axis of [010] direction \cite{TSY+88, YTTS89, BCF+91} is achieved below $T_{N1}=212$ K with the superexchange coupling of $J=67{\pm}20$ meV \cite{YTS88,YTTS89,BCF+91} which is an order of magnitude larger than that expected from the N\'{e}el temperature, suggesting a strongly correlated and low dimensional spin system \cite{IIS+90, ZBC+90, GRW91}, (3) the significantly reduced Cu$^{2+}$ spin moment $0.65 \mu_{B}$ compared with $1 \mu_{B}$ expected for a Cu$^{2+}$ ion is observed \cite{YTS88,YTTS89,LMJ+89,BCF+91} at $T=4$ K due either to quantum spin fluctuations in low dimensional system or to covalent effect, (4) the temperature dependence of paramagnetic susceptibility \cite{OS62,RBF87,KOS+88, KC91,CRS94} shows a broad peak at around 540 K \cite{OS62}, reminiscent of quasi 1D or 2D antiferromagnet, and (5) CuO belongs to the charge-transfer gap insulators \cite{ETS90}.  

All the $3d$ transition metal monoxides, with the only exception of CuO, are 3D antiferromagnets.  
It may be unexpected to find a low dimensional magnetism in such a chemically 3D structure as the monoxide CuO.  
In fact, many experimental data of CuO have been explained in the scheme of quasi 1D antiferromagnet.  
It has been assumed that the strongest antiferromagnetic coupling may reside on a particular Cu-O-Cu bond \cite{OS62,YTTS89} by invoking the Anderson model of superexchange interactions which tells the larger bond angle preferred for the stronger  antiferromagnetic coupling \cite{And64,HTH75}.  

The proposed picture of spin 1D chain, however, would not be obvious, because the intra-chain bond is even longer than the inter-chain ones.  
We note a close competition present among the bond angles as well as bond lengths in CuO.   The 1D picture may be conceivable but still only a speculation unless quantitative evidence for the bond angle scheme is presented.

We will find below that making a distinction between 1D and 2D involves a delicate problem.   No investigation, to our knowledge, has been performed yet to make a comparison between 1D and 2D in CuO.  
The following questions are still open: (1) Which is the better model to describe the paramagnetic state of CuO, 1D or 2D? (2) If it turns out to be a 1D antiferromagnet as has been believed so far, what is the cause for a particular bond to be magnetically active and others less active?

In this paper, we present evidence for a quasi 1D antiferromagnetism in CuO beyond 2D antiferromagnetism.  
We carried out the $^{63}$Cu-NMR shift measurement to obtain an  absolute value of the spin susceptibility which provides us an opportunity to make direct comparison with theoretical predictions.  
The theories we refer to in the present work are those studied in a framework of the Heisenberg model.  
Although the earlier studies about susceptibility have shown an experimental indication being consistent with the prediction of 1D AFH, a lack of the knowledge about absolute value of spin susceptibility prevents quantitative comparison between experiment and theory.   
We will find below that the absolute value is crucial to make a distinction between the 1D and 2D antiferromagnet.  

A susceptibility measurement gives us only a total susceptibility which consists of the three contributions from the spin, orbital and ionic-core.  
Only the spin contribution $\chi_{spin}$ is required to be compared with a quantum spin theory.  
A direct measurement of spin contribution can in principle be made by neutron experiment, but the measurement would be difficult because of poor signal intensity in case of weak paramagnetism like CuO.  
Ionic-core contribution $\chi_{core}$ can be estimated from the ionic data-base or theoretical calculation.  
Orbital contribution $\chi_{orb}$ arises from second-order excitations between crystal field levels, being material dependent and thus to be measured by an experiment like NMR.
  
In the absence of experimental information of orbital contribution, as is common, its contribution would be treated as an adjustable parameter to make a qualitative comparison in the temperature dependence of spin contribution between experiment and theory.  
The aim of our study is to present the experimental absolute values of the orbital and spin contributions in CuO to make a quantitative comparison possible.  
\begin{figure}
\includegraphics{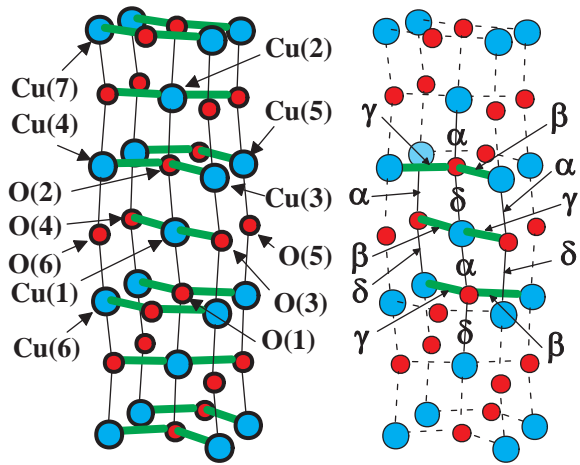}
\caption{\label{fig:crystal}Crystal structure of CuO.  
The big and small balls are copper and oxygen ions, respectively.  
The left panel: We define Cu(1) through Cu(7) for the convenience of discussion, although  all the Cu sites are crystallographically equivalent.  
Similarly, oxygen ions, which occupy the other equivalent sites, are labeled as O(1) through O(6).  
The primitive directions of [100], [010] and [001] are along lines connecting Cu(3) to Cu(6), Cu(5) to Cu(3), and Cu(3) to Cu(7), respectively.  
The thin bonds running vertically are along the chain direction [10$-$1], the fat bonds connect the chains.  
The right panel: There are four bond lengths in the crystal.  
We call them $\alpha$, $\beta$, $\gamma$ and $\delta$ in the order of short.  
The values of them are given in Table~\ref{tab:crystal1}.  
The intra-chain bonds consist of the alternative  sequence of $\alpha$ and $\delta$,  the inter-chain bonds similarly of $\beta$ and $\gamma$.}
\end{figure}

\section{Crystal Structure and Sample Preparation}
CuO crystallizes in a monoclinic structure with a space group of $Cc$ \cite{AW91}.  
The crystal structure, shown in Fig.~\ref{fig:crystal} and Table~\ref{tab:crystal1}, is a unique one among the transition metal monoxides, rather than the cubic rock salt structure of the most.  The unit cell contains four copper ions which are all crystallographically equivalent, as well as four equivalent oxygens.   

\begin{figure}
\includegraphics{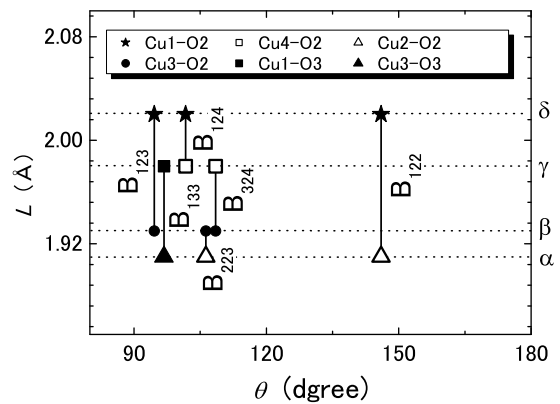}
\caption{\label{fig:bond}The bond length $L$ is plotted against the bond angle $\theta$.  The vertical bars labeled by B$_{ijk}$ denote the bonds Cu($i$)-O($j$)-Cu($k$) as defined in Table~\ref{tab:crystal2}.}
\end{figure}

A significant feature can be seen from Table~\ref{tab:crystal2} and Fig.~\ref{fig:bond}  that the bond Cu(1)-O(2)-Cu(2) (abbreviated as B$_{122}$ in the following) has a noticeably larger bond angle compared with all others.  
This is the reason why this bond has been assumed to carry the strongest superexchange interaction \cite{OS62, YTTS89}.  
This assumption seems conceivable, but still only a speculation which needs evidence to specify the bond angle and bond length dependences of the superexchange interaction.  

\begin{table}
\caption{\label{tab:crystal1}The bond lengths (inter-atomic distances) are shown in the units of \AA.  We denote the four shortest bonds as
; $\alpha=1.91$ \AA, $\beta=1.93$ \AA, $\gamma=1.98$ {\AA}  and $\delta=2.02$ \AA.}
\begin{ruledtabular}
\begin{tabular}{ccccccccc}
      & Cu(1) & Cu(2) & O(1) & O(2) & O(3) & O(4) & O(5) & O(6) \\
\hline
Cu(1) &		&		& $\alpha$ & $\delta$ & $\gamma$ & $\beta$ & 2.79 & 2.78 \\
Cu(2) & 3.76 & 		&		  &  $\alpha$ & 		  &		 	&	  & \\
Cu(3) & 2.90 & 3.07   & 		  & $\beta$   & $\alpha$ & & & \\
Cu(4) & 3.10 & 2.90	&	& $\gamma$ & & & & \\
\end{tabular}
\end{ruledtabular}
\end{table}
\begin{table}
\caption{\label{tab:crystal2}The bond length $L$ (measured along the bond Cu-O-Cu) and bond angle $\theta$ are shown.  There are six kinds of Cu-O-Cu bond in the crystal.  We define the abbreviation B$_{ijk}$ representing the bond Cu($i$)-O($j$)-Cu($k$).
}
\begin{ruledtabular}
\begin{tabular}{lccc}
      & \multicolumn{2}{c}{$L$} & $\theta$  \\
& & \AA & degree \\
\hline
B$_{123}$ &	$\beta+\delta$	& 3.95 & 94.6 \\
B$_{133}$ & $\alpha+\gamma$ & 3.89	&	96.8 \\
B$_{124}$ & $\gamma+\delta$ & 4.00 & 101.7 \\
B$_{223}$ & $\alpha+\delta$ & 3.84	& 106.3 \\
B$_{324}$ & $\beta+\gamma$ & 3.91 & 108.5 \\
B$_{122}$ & $\alpha+\delta$ & 3.93 & 146.0 \\
\end{tabular}
\end{ruledtabular}
\end{table}

We began our sample preparation by initializing the stoichiometry of commercially obtained CuO, in which the powder sample of the nominal CuO was annealed at 540 $^{\circ}$C in air for a day.  
We have found that the annealing process is crucial to have the Cu NQR/NMR signal visible at any temperatures including 4 K and 300 K, provably because a possible oxygen deficiency may be removed by the annealing process.  

The annealed sample was pulverized to be fine powder with 5 $\mu$m in typical diameter, and then to be mixed with epoxy resin and finally exposed in a magnetic field (7 T) to prepare a magnetically aligned sample.  
The X-ray diffraction data on the aligned sample indicates the [010] direction preferably oriented along the magnetic field.  
The degree of alignment is estimated to be $42\pm10$ {\%} by a method described later.  
The other crystal axis directions may be randomly oriented in the plane perpendicular to the magnetic field.
\begin{figure}
\includegraphics{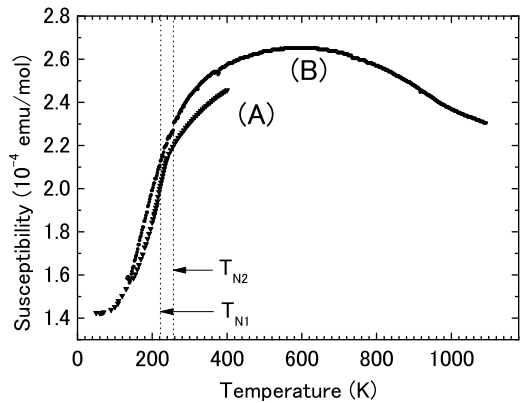}
\caption{\label{fig:chipoly}Polycrystal susceptibility of CuO .  
The present sample (A) and literature data (B) taken from Ref.~\onlinecite{OS62} are shown.  
The arrows indicated by $T_{N1}$ (212 K) and $T_{N2}$ (231 K) are the 1st and 2nd N\'{e}el temperature, respectively}
\end{figure}
\begin{figure}
\includegraphics{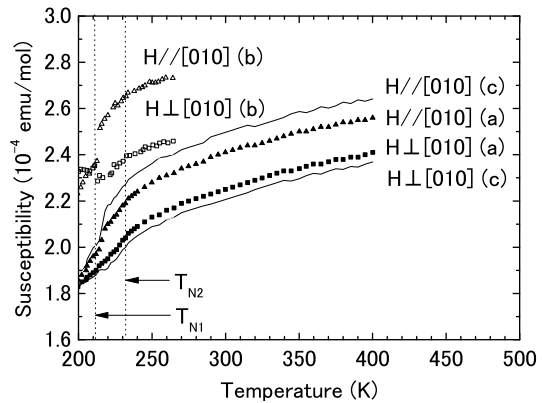}
\caption{\label{fig:chianiso}Anisotropic susceptibility of CuO.  
The data set of (a) is the magnetically [010] aligned polycrystal susceptibility by the present work, the data set of (b) is the single crystal susceptibility taken form Ref.~\onlinecite{KC91}, and the data set of (c) is given by an alignment correction to the data set of (a) by the method described in the text.  
The origin of the offset found in the absolute value of (b) is unknown, although we have observed a similar offset in a single crystal prepared in our Institute (Ref.~\onlinecite{CKM+97}).  }
\end{figure}
\section{Susceptibility and NMR Measurements}
The susceptibility measurement has been made by a SQUID magnetometer with an applied magnetic field of 4 T.  
Fig.~\ref{fig:chipoly} shows the polycrystal susceptibility of the epoxy-free sample.  
Among the literatures of susceptibility measurement, Ref.~\onlinecite{OS62} is the only work which has reported the high temperature data beyond the susceptibility maximum.  
The small discrepancy found between (A) and (B) is unknown.  
We will use below the data of Ref.~\onlinecite{OS62} to make a comparison in the isotropic part of spin susceptibility between the experiment and theory.

We need an anisotropic susceptibility data to get an orbital susceptibility which is primarily anisotropic.  
The uniaxially anisotropic susceptibility of the present magnetically-aligned sample has been obtained as is shown in Fig.~\ref{fig:chianiso}.  
Single crystal data by Ref.~\onlinecite{KC91} shows a larger anisotropy than the present sample, provably because the present magnetic alignment is only partial.  
The temperature range of the single crystal data by Ref.~\onlinecite{KC91}, unfortunately, is limited below 264 K, too narrow to meet the following analysis together with NMR data we measured.  
Thus we use two types of susceptibility data in the following analysis to get the orbital susceptibility; one is the raw data (a) as is measured, another is the corrected data (c) given by a method described below.  
It can be seen later that the main conclusion drawn from the following analysis turns out to be independent of which susceptibility data are used.

The corrected data have been taken as follows.  
We assume the present sample is a sum of crystallites oriented perfectly and polycrystalites oriented randomly, then we can write, 
\begin{subequations}
\label{eq:chianiso}
\begin{equation}
\chi_{p}^{\parallel}=p\chi_{s}^{\parallel} + \frac{1}{3}(1-p)(\chi_{s}^{\parallel}+2\chi_{s}^{\perp}),
\label{subeq:chiparallel}
\end{equation}
\begin{equation}
\chi_{p}^{\perp}=p\chi_{s}^{\perp} + \frac{1}{3}(1-p)(\chi_{s}^{\parallel}+2\chi_{s}^{\perp}),
\label{subeq:chiperp}
\end{equation}
\end{subequations}
where $p$ is the degree of alignment, $\chi_{p}$ and $\chi_{s}$ are the susceptibilities of  partially aligned sample and single crystal one, respectively,  measured in a field parallel ($\chi^{\parallel}$) and perpendicular ($\chi^{\perp}$) to the [010] direction.
Comparing the anisotropy of the raw data (a) with that of the single crystal data (b) at 264 K gives $p=0.42\pm10$.  
The degree of alignment estimated in this manner is found nearly temperature independent in the paramagnetic region, as is expected.  
We can estimate the fully aligned susceptibility in the temperature range of $231-400$ K by substituting the $p$ and raw data to the inverse of Eqs.~(\ref{eq:chianiso}).
The result is shown in Fig.~\ref{fig:chianiso} (c).
\begin{figure}
\includegraphics{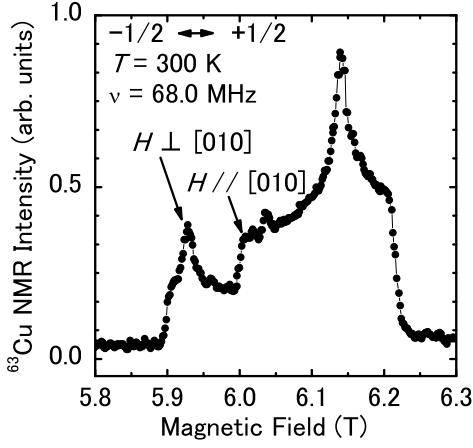}
\caption{\label{fig:spectrum}$^{63}$Cu-NMR $-1/2 \leftrightarrow +1/2$ spectrum taken at 300 K and 68.0 MHz in the present sample (not aligned).  
The spectrum shows a powder pattern which undergoes the 2nd order quadrupole effect as well as magnetic shift.  
Each point has been taken by accumulating about a thousand of spin echo signals.  
The signal has become hard to be observed below 240 K, because of the critical slowing dawn of spin fluctuations towards the magnetic phase transition at 231 K.  
The arrows indicate the particular resonance positions of which we took the frequency dependences to deduce the NMR shifts $K$.  }
\end{figure}
\begin{figure}
\includegraphics{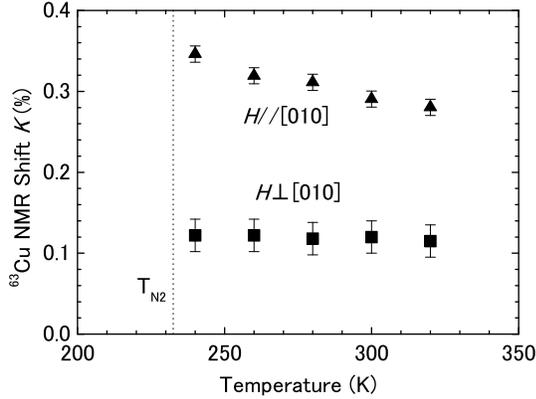}
\caption{\label{fig:K-T}The Cu-NMR paramagnetic shift with respect to the effective gyromagnetic ratio of $^{63}\gamma/(2\pi)=11.284$ MHz/T is plotted against temperature.  
We measured the anisotropic shift for the two particular field directions parallel and perpendicular to [010] axis at each temperature.  
The $T_{N2}$ (231 K) indicates the 2nd N\'{e}el temperature. }
\end{figure}
\begin{figure}
\includegraphics{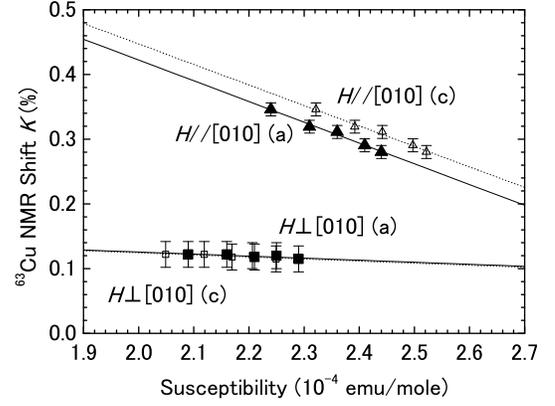}
\caption{\label{fig:K-chi}The Cu-NMR shift is plotted against susceptibility with the temperature as implicit parameter.  The data set of closed symbols (a) are given from the susceptibility raw data (a) of Fig.~\ref{fig:chipoly}, the data set of open symbols (c) are from the corrected susceptibility data (c) shown in Fig.~\ref{fig:chipoly}, respectively.  The solid and dotted lines are linear fits through the respective data sets. }
\end{figure}

Paramagnetic $^{63}$Cu-NMR shift $K$ is obtained by measuring a frequency dependence of the spectrum (Fig.~\ref{fig:spectrum}), and the result is plotted in Fig.~\ref{fig:K-T}. 
In order to deduce the magnetic contribution of $K$ separately from the quadrupolar one in the spectrum, we have eliminated the quadrupole effect \cite{CBK} by taking a frequency dependence of the particular resonance positions indicated by arrows in Fig.\ref{fig:spectrum}.  

The orbital susceptibility is deduced, as is listed in Tab.~\ref{tab:chiorbital}, by a well-known method \cite {CJY64} of analyzing the $K$-$\chi$ plot diagram (Fig.~\ref{fig:K-chi}) together with the following relations between the total shift $K$ and total susceptibility $\chi$ written as, 
\begin{subequations}
\label{eq:Kandchi}
\begin{equation}
K=K_{spin}(T) + K_{orb},
\label{subeq:Ktotal}
\end{equation}
\begin{equation}
\chi=\chi_{spin}(T) + \chi_{orb} + \chi_{core},
\label{subeq:chitotal}
\end{equation}
\end{subequations}
and
\begin{subequations}
\label{eq:K-chi}
\begin{equation}
K_{spin}(T)={\frac{A_{spin}}{N_A \mu_{B}}}{\chi_{spin}(T)},
\label{subeq:Kspin}
\end{equation}
\begin{equation}
K_{orb}={\frac{2\langle r^{-3}\rangle \xi^2}{N_A}}{\chi_{orb}},
\label{subseq:Korb}
\end{equation}
\end{subequations}
where we use the usual notations as the hyperfine coupling tensor $A_{spin}$ including the Fermi contact and dipole interaction, the Avogadro's number $N_A$, the Bohr magneton ${\mu_B}$, expectation value $\langle r^{-3}\rangle$ within the $3d^9$ radial wave function and the covalence reduction factor $0\leq \xi^2 \leq 1$. 

The orbital susceptibility $\chi_{orb}$ tells us the $g$ factor, as listed in Table~\ref{tab:chiorbital}, by using the following relations as \cite{AB}, 
\begin{subequations}
\label{eq:chiorb}
\begin{equation}
\chi_{orb}^{\parallel}=\frac{8N_A {\mu_B}^2 {\xi}^2}{{\Delta}_{0}},
\label{subeq:chiorbparallel}
\end{equation}
\begin{equation}
\chi_{orb}^{\perp}=\frac{2N_A {\mu_B}^2 \xi^2}{\Delta_1},
\label{subeq:chiorbperp}
\end{equation}
\end{subequations}
and
\begin{subequations}
\label{eq:g}
\begin{equation}
g^{\parallel}=2-\frac{8\lambda_{so}{\xi}^2}{{\Delta}_{0}},
\label{subeq:gparallel}
\end{equation}
\begin{equation}
g^{\perp}=2-\frac{2\lambda_{so}{\xi}^2}{{\Delta}_{1}},
\label{subeq:gperp}
\end{equation}
\end{subequations}
where we denote  the spin-orbit coupling $\lambda_{so}$ of the $3d^9$ state, and the crystal field splittings $\Delta_0$ and $\Delta_1$ of Cu$^{2+}$ ions in a tetragonal crystal field.  

In Table~\ref{tab:chiorbital} we assume the following values, as commonly expected for cuprates \cite{AB}, that the total ionic core susceptibility of Cu$^{2+}$ and O$^{2-}$ is $-0.23{\times}10^{-4}$ emu/mol \cite{LB}, the hyperfine radius parameter $\langle r^{-3} \rangle=8.25$ a.u., covalence reduction factor $\xi^2=0.75$ \cite{Shi93} and spin-orbit coupling  $\lambda_{so}=-0.095$ eV \cite{SAY+93}. 

\begin{table}
\caption{\label{tab:chiorbital}The orbital susceptibility $\chi_{orb}$, crystal field splitting $\Delta$, and $g$ factor are obtained by analyzing the straight lines shown in Fig.~\ref{fig:K-chi} together with Eqs.~\ref{eq:Kandchi}-\ref{eq:g}.  
The column (a) and (c) are obtained from the data sets of (a) and (c) in Fig.~\ref{fig:K-chi}, respectively.   The susceptibility and crystal field splitting are in the units of $10^{-4}$ emu/mol and eV, respectively.
}
\begin{ruledtabular}
\begin{tabular}{ccccccc}
& $\chi_{orb}^{\parallel}$ & $\chi_{orb}^{\perp}$ & $\Delta_0$ & $\Delta_1$ & $g^{\parallel}$ & $g^{\perp}$ \\
\hline
(a) & 0.672 & 0.139 & 2.26 & 2.72 & 2.19 & 2.04 \\
(c) & 0.674 & 0.139 & 2.22 & 2.72 & 2.2 & 2.04 \\
\end{tabular}
\end{ruledtabular}
\end{table}
\section{Spin Susceptibility and Superexchange Interaction}
The experimental spin susceptibility $\chi_{spin}$ can be obtained by subtracting the orbital and ionic core susceptibilities from the observed total susceptibility.  
For the total susceptibility to be subtracted, we use here the data of Ref.~\onlinecite{OS62} rather than the present data, in order to show the whole behavior including the susceptibility maximum.  
We take the isotropic component of orbital susceptibility from Table~\ref{tab:chiorbital} as $\chi_{orb}^{iso}=\frac{1}{3}(\chi_{orb}^{\parallel} + 2\chi_{orb}^{\perp})=0.31\times 10^{-4} $emu/mol and the literature value of $\chi_{core}=-0.23\times 10^{-4}$emu/mol \cite{LB} for the subtraction.  
We note here that the difference in the orbital susceptibility between the column (a) and (c) in Table~\ref{tab:chiorbital} is about 3 \% of the observed total susceptibility, being small enough to make only a negligible difference in the result of spin susceptibility.
This is also related with the principle that the spin susceptibility of a $S=1/2$ system is primarily isotropic.  

\begin{figure}
\includegraphics{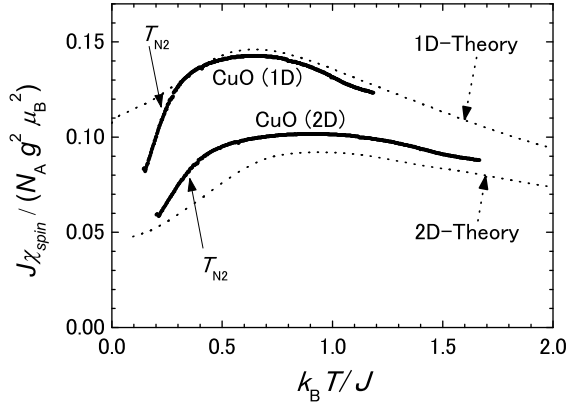}
\caption{\label{fig:chispin}Comparison between experimental and theoretical spin susceptibility.  The dotted curves are theoretical predictions for 1D \cite{EAT94} and 2D \cite{OKN88} $S=1/2$ Heisenberg antiferromagnets.  
We get the experimental curves ``1D'' and ``2D'' by applying $J= 73 \pm 3 $ (1D) and 52 $\pm 3$ meV (2D), respectively, to the experimental spin susceptibility and temperature.}
\end{figure}

In Fig.~\ref{fig:chispin}, the dotted curves are theoretical predictions for 1D \cite{EAT94} and 2D \cite{OKN88} $S=1/2$ Heisenberg antiferromagnet.  Since the theoretical predictions are given in a scale normalized by the coupling constant $J$ for both temperature and susceptibility axes, we need a particular value of $J$ to make a comparison possible between experiment and theory.  
Each theory predicts a relation between $J$ and $T_{max}$ given by,
\begin{equation}
 J=k_{B}T_{max}/d,
\label{eq:J-Tmax}
\end{equation}
where $T_{max}$ is the temperature showing susceptibility maximum, $d$ is the numerical factor given by $d=$0.64 and 0.9 for 1D and 2D, respectively, and $J$ is defined in a spin Hamiltonian expressed by,
\begin{equation}
\mathcal{H}=-J S_1 S_2.
\label{eq:Hamiltonian}
\end{equation}

Putting the experimental value $T_{max}=540 \pm 20$ K \cite{OS62} into Eq.~\ref{eq:J-Tmax}, we get $J=73 \pm 3$ and $52 \pm 2$ meV for 1D and 2D, respectively, and correspondingly the experimental curves ``1D'' and ``2D'' , respectively, as shown in Fig.~\ref{fig:chispin}.  
The $g$ factor used in the normalized susceptibility axis is given by substituting the values listed in Tab.~\ref{tab:chiorbital} into the isotropic component expressed as $g=\frac{1}{3}(g^{\parallel} + 2g^{\perp})=2.09$.  

We can find from Fig.~\ref{fig:chispin} that the 1D theory reproduces nicely the experimental 1D curve above the N\'{e}el temperature $T_{N2}$.  
The 2D case shows a poorer agreement between the 2D theory and the experimental 2D curve at all temperatures.  
This gives a support for the hypothesis that a quasi 1D AFH model is a good approximation of CuO.  

We emphasize here that the discrepancy between the 2D theory and the experimental 2D curve  is near the isotropic component of orbital susceptibility, which suggests that estimating orbital susceptibility is crucial to make a comparison in the absolute value of spin susceptibility.  
If the orbital susceptibility is taken as an adjustable parameter, as is often assumed, it would be hard to make a distinction between 1D and 2D spin susceptibility.  
\begin{table}
\caption{\label{tab:J-bond}The geometrical parameters of Cu-O-Cu bond (bond angle $\theta$ and bond length $L$) and superexchange coupling $J$ in the antiferromagnetic cuprates are listed.  
The $J$ is defined by Eq.~\ref{eq:Hamiltonian} in the text or by the 1st nearest superexchange coupling in the case of CuGeO$_3$.  
The data of $J$ are taken from experiments of magnetic susceptibility, neutron and Raman scattering measurements. 
The bond angles and bond lengths are obtained by calculations using the literature data about the crystal structures; CuGeO$_3$ from Ref.~\onlinecite{HHY+97}; CuO from Ref.~\onlinecite{AW91}; YBa$_2$Cu$_3$O$_6$, La$_2$CuO$_4$ and Nd$_2$Cu$_4$ from Ref.~\onlinecite{Pic89}.  
The $\theta$ and $L$ of CuO refer to the bond B$_{122}$. 
 }
\begin{ruledtabular}
\begin{tabular}{lccccc}
& $\theta$ & $L$  & \multicolumn{3}{c}{$J$ (meV)} \\
& (degree) & (\AA) & Suscept. & Neutron & Raman \\
\hline
CuGeO$_3$ & 98.4 & 3.884 & $6.9\pm3$\footnotemark[1] & $10.4$\footnotemark[2] & \\
CuO & 145.9 & 3.926 & $72.8\pm3$\footnotemark[3] & $67\pm20$\footnotemark[4] & \\
YBa$_2$Cu$_3$O$_6$ & 166.9 & 3.882 &  & $120\pm20$\footnotemark[5] & 120\footnotemark[6] \\
La$_2$CuO$_4$ & 173.2 & 3.810 & $136$\footnotemark[7] & $133\pm3$\footnotemark[8] & $137$\footnotemark[9] \\
Nd$_2$Cu$_4$ & 180 & 3.908 &  & $155\pm3$\footnotemark[8] &\\
\end{tabular}
\end{ruledtabular}
\footnotetext[1]{Ref.~\onlinecite{FKL+98}}
\footnotetext[2]{Ref.~\onlinecite{NFA94}}
\footnotetext[3]{this work}
\footnotetext[4]{Ref.~\onlinecite{YTTS89}}
\footnotetext[5]{Ref.~\onlinecite{SST+93}}
\footnotetext[6]{Ref.~\onlinecite{LFSW88}}
\footnotetext[7]{Ref.~\onlinecite{Joh89}} 
\footnotetext[8]{Ref.~\onlinecite{BCIP98}}
\footnotetext[9]{Ref.~\onlinecite{LFR+88}}
\end{table}
\begin{figure}
\includegraphics{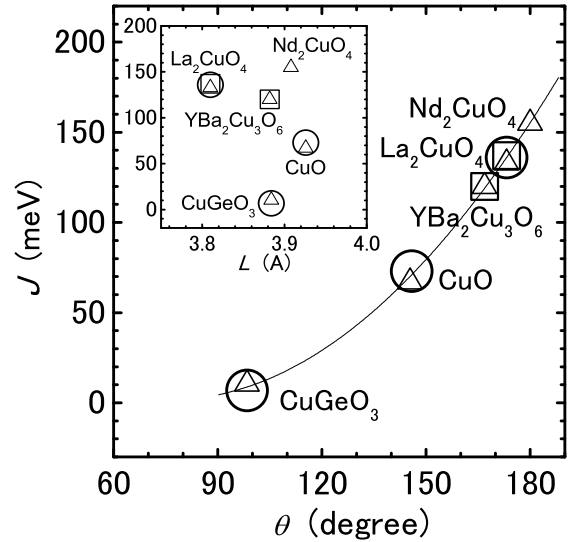}
\caption{\label{fig:J-bond}The superexchange coupling $J$ is plotted against the bond angle $\theta$ and against the bond length $L$ (the inset) in the cuprates.    
The data are taken form Table~\ref{tab:J-bond}.  
The solid curve shows a cubic fit through the points.  
The triangles are the data taken from neutron scattering, the squares from Raman scattering, the circles from susceptibility measurements. }
\end{figure}

The earlier result of $J=69 \pm 20 $ meV by neutron measurement \cite{YTTS89} is in better agreement with the present value of $73 \pm 3$ meV (1D), than that of 52$\pm2$ meV (2D).  
This yields another support for the quasi 1D antiferromagnetism of CuO.  
The present agreement may be sufficient to argue that the effect of 1D quantum spin fluctuations \cite{EAT94} is responsible for the reduction of the spin moment 0.65 $\mu_B$ compared with 1 $\mu_B$ expected for a Cu$^{2+}$  ion.  

The chain direction of [10$-$1] may be most promising from the viewpoint of the bond angle of the bond B$_{122}$.  
The advantage by the bond angle of B$_{122}$ would, nevertheless, trade off the disadvantage by the bond length of itself.  
A careful explanation for the roles of bond angle and bond length in magnetic coupling $J$ is required.  
We will be focused to this point in the followings.

Table~\ref{tab:J-bond} summarizes the typical examples of $J$ obtained experimentally for the series of cuprates, together with the bond angle and bond length.  
We plot them in Fig.~\ref{fig:J-bond}, in which a systematic correlation is found between $J$ and $\theta$, but no straightforward correlation between $J$ and $L$.  
The $\theta$ and $L$ of CuO refer to the bond B$_{122}$.  
If we take the inter-chain bonds B$_{ijk}$ rather than B$_{122}$, no correlation was found in both the plots.  

It can, in principle, be expected that the superexchange coupling $J$ would be given by a function of both bond angle and bond length.
Fig.~\ref{fig:J-bond} implies that the bond angle is effectively of a prime importance in the magnetic coupling of cuprates, and the variable range allowed for bond length in the  cuprates may be narrower than that of bond angle, the latter being capable of a full range from 90 degree through 180 degree.

A bond angle dependence of $J$ in the ferromagnetic cuprates has also been found in Ref.~\onlinecite{MTM+98} in which they have reproduced theoretically the bond angle dependence of the ferromagnetic $J$ in the vicinity of $\theta=90$ degree by a close inspection into  the contribution from the charge-transfer gap $\Delta$ to the superexchange coupling $J$.
This tempts us to believe that the bond angle dependence of antiferromagnetic $J$ found in the present study is conceivable.  

We can find from Fig.~\ref{fig:J-bond} a reason why CuO is a quasi 1D antiferromagnet.
The $\theta$ dependence of $J$ tells us that all the inter-chain bonds Cu-O-Cu, having the bond angles ranging from 94.6 to 108.5 degree (see Table~\ref{tab:crystal2}), can be expected to have the $J$ values smaller than 15 meV.  
The ratios of the inter-chain $J$ values (15 meV or less) with the intra-chain $J$ value (73 meV) are 0.2 or less, which may be sufficient to make CuO a quasi 1D antiferromagnet.

\section{Discussion}

We have assumed during the present study that the Heisenberg model is a good approximation for the spin system.  
It could be a possible issue to examine to what extent the 2nd nearest neighbor Cu$^{2+}$ ion comes into magnetic interaction.  
This may be investigated in a future work, but we can note at the present time that a possible 2nd nearest superexchange coupling in the spin chain seems negligible, as far as the susceptibility is concerned.  

This gives a contrast to the case of the 1D antiferromagnet CuGeO$_3$ where the spin susceptibility is considerably reduced, compared to the 1D Heisenberg model, provably by the Majumdar-Ghosh type of a spin-gap due to the significant contribution from the 2nd nearest superexchange interaction in the chain \cite{YS97, FKL+98}.  
The crystal structure may be responsible for the difference between CuO and CuGeO$_3$. 
The latter has a edge sharing spin chain in which a possible overlap integral between the two oxygens locating along the chain may increase the 2nd nearest superexchange interaction.   

We do not think the bond angle dependence shown in Fig.~\ref{fig:J-bond} to be an absolute relation, but expect it to be an empirical relation valid in some cases.  
For example, considering the mirror and inversion symmetry of the Cu-O-Cu bond, the bond angle dependence of $J$ would have a zero gradient at $\theta=180$ degree, being not consistent with the present result in Fig.~\ref{fig:J-bond}.  
A broken-symmetry caused by the surrounding ions such as alkaline-earth and rare-earth metals in YBa$_2$Cu$_3$O$_6$ and La$_2$CuO$_4$, which are located at sites without mirror symmetry on the CuO$_2$ layer, may be responsible for the singular behavior found near $\theta=180$ degree.  

One of the authors (TS) expresses his thanks to Masashi Hase, Sadamichi Maekawa, Russell E. Walstedt, Noriaki Hamada, Takami Tohyama, Hisatoshi Yokoyama and Masayuki Itoh for their valuable discussions.  
%


\begin{thebibliography}{38}
\expandafter\ifx\csname natexlab\endcsname\relax\def\natexlab#1{#1}\fi
\expandafter\ifx\csname bibnamefont\endcsname\relax
  \def\bibnamefont#1{#1}\fi
\expandafter\ifx\csname bibfnamefont\endcsname\relax
  \def\bibfnamefont#1{#1}\fi
\expandafter\ifx\csname citenamefont\endcsname\relax
  \def\citenamefont#1{#1}\fi
\expandafter\ifx\csname url\endcsname\relax
  \def\url#1{\texttt{#1}}\fi
\expandafter\ifx\csname urlprefix\endcsname\relax\def\urlprefix{URL }\fi
\providecommand{\bibinfo}[2]{#2}
\providecommand{\eprint}[2][]{\url{#2}}

\bibitem[{\citenamefont{Forsyth et~al.}(1988)\citenamefont{Forsyth, Brown, and
  Wanklyn}}]{FBW88}
\bibinfo{author}{\bibfnamefont{J.~B.} \bibnamefont{Forsyth}},
  \bibinfo{author}{\bibfnamefont{P.~J.} \bibnamefont{Brown}}, \bibnamefont{and}
  \bibinfo{author}{\bibfnamefont{B.~M.} \bibnamefont{Wanklyn}},
  \bibinfo{journal}{J. Phys. C} \textbf{\bibinfo{volume}{21}},
  \bibinfo{pages}{2917} (\bibinfo{year}{1988}).

\bibitem[{\citenamefont{Yang et~al.}(1989)\citenamefont{Yang, Thurston,
  Tranquada, and Shirane}}]{YTTS89}
\bibinfo{author}{\bibfnamefont{B.~X.} \bibnamefont{Yang}},
  \bibinfo{author}{\bibfnamefont{T.~R.} \bibnamefont{Thurston}},
  \bibinfo{author}{\bibfnamefont{J.~M.} \bibnamefont{Tranquada}},
  \bibnamefont{and} \bibinfo{author}{\bibfnamefont{G.}~\bibnamefont{Shirane}},
  \bibinfo{journal}{Phys. Rev. B} \textbf{\bibinfo{volume}{39}},
  \bibinfo{pages}{4343} (\bibinfo{year}{1989}).

\bibitem[{\citenamefont{Loram et~al.}(1989)\citenamefont{Loram, Mirza, Joyce,
  and Osborne}}]{LMJ+89}
\bibinfo{author}{\bibfnamefont{J.~W.} \bibnamefont{Loram}},
  \bibinfo{author}{\bibfnamefont{K.~A.} \bibnamefont{Mirza}},
  \bibinfo{author}{\bibfnamefont{C.~P.} \bibnamefont{Joyce}}, \bibnamefont{and}
  \bibinfo{author}{\bibfnamefont{A.~J.} \bibnamefont{Osborne}},
  \bibinfo{journal}{Europhys. Lett.} \textbf{\bibinfo{volume}{8}},
  \bibinfo{pages}{263} (\bibinfo{year}{1989}).

\bibitem[{\citenamefont{Tsuda et~al.}(1988)\citenamefont{Tsuda, Shimizu,
  Yasuoka, Kishio, and Kitazawa}}]{TSY+88}
\bibinfo{author}{\bibfnamefont{T.}~\bibnamefont{Tsuda}},
  \bibinfo{author}{\bibfnamefont{T.}~\bibnamefont{Shimizu}},
  \bibinfo{author}{\bibfnamefont{H.}~\bibnamefont{Yasuoka}},
  \bibinfo{author}{\bibfnamefont{K.}~\bibnamefont{Kishio}}, \bibnamefont{and}
  \bibinfo{author}{\bibfnamefont{K.}~\bibnamefont{Kitazawa}},
  \bibinfo{journal}{J. Phys. Soc. Jpn.} \textbf{\bibinfo{volume}{57}},
  \bibinfo{pages}{2908} (\bibinfo{year}{1988}).

\bibitem[{\citenamefont{Brown et~al.}(1991)\citenamefont{Brown, Chattopadhyay,
  Forsyth, and Nunez}}]{BCF+91}
\bibinfo{author}{\bibfnamefont{P.~J.} \bibnamefont{Brown}},
  \bibinfo{author}{\bibfnamefont{T.}~\bibnamefont{Chattopadhyay}},
  \bibinfo{author}{\bibfnamefont{J.~B.} \bibnamefont{Forsyth}},
  \bibnamefont{and} \bibinfo{author}{\bibfnamefont{V.}~\bibnamefont{Nunez}},
  \bibinfo{journal}{J. Phys. Conds. Matter} \textbf{\bibinfo{volume}{3}},
  \bibinfo{pages}{4281} (\bibinfo{year}{1991}).

\bibitem[{\citenamefont{Yang et~al.}(1988)\citenamefont{Yang, Thurston, and
  Shirane}}]{YTS88}
\bibinfo{author}{\bibfnamefont{B.~X.} \bibnamefont{Yang}},
  \bibinfo{author}{\bibfnamefont{T.~R.} \bibnamefont{Thurston}},
  \bibnamefont{and} \bibinfo{author}{\bibfnamefont{G.}~\bibnamefont{Shirane}},
  \bibinfo{journal}{Phys. Rev. B} \textbf{\bibinfo{volume}{38}},
  \bibinfo{pages}{174} (\bibinfo{year}{1988}).

\bibitem[{\citenamefont{Itoh et~al.}(1990)\citenamefont{Itoh, Imai, Shimizu,
  Tsuda, Yasuoka, and Ueda}}]{IIS+90}
\bibinfo{author}{\bibfnamefont{Y.}~\bibnamefont{Itoh}},
  \bibinfo{author}{\bibfnamefont{T.}~\bibnamefont{Imai}},
  \bibinfo{author}{\bibfnamefont{T.}~\bibnamefont{Shimizu}},
  \bibinfo{author}{\bibfnamefont{T.}~\bibnamefont{Tsuda}},
  \bibinfo{author}{\bibfnamefont{H.}~\bibnamefont{Yasuoka}}, \bibnamefont{and}
  \bibinfo{author}{\bibfnamefont{Y.}~\bibnamefont{Ueda}}, \bibinfo{journal}{J.
  Phys. Soc. Jpn.} \textbf{\bibinfo{volume}{59}}, \bibinfo{pages}{1143}
  (\bibinfo{year}{1990}).

\bibitem[{\citenamefont{Ziolo et~al.}(1990)\citenamefont{Ziolo, Borsa, Corti,
  Rigamonti, and Parmigiani}}]{ZBC+90}
\bibinfo{author}{\bibfnamefont{J.}~\bibnamefont{Ziolo}},
  \bibinfo{author}{\bibfnamefont{F.}~\bibnamefont{Borsa}},
  \bibinfo{author}{\bibfnamefont{M.}~\bibnamefont{Corti}},
  \bibinfo{author}{\bibfnamefont{A.}~\bibnamefont{Rigamonti}},
  \bibnamefont{and}
  \bibinfo{author}{\bibfnamefont{F.}~\bibnamefont{Parmigiani}},
  \bibinfo{journal}{J. Appl. Phys.} \textbf{\bibinfo{volume}{67}},
  \bibinfo{pages}{5864} (\bibinfo{year}{1990}).

\bibitem[{\citenamefont{Graham et~al.}(1991)\citenamefont{Graham, Riedi, and
  Wanklyn}}]{GRW91}
\bibinfo{author}{\bibfnamefont{R.~G.} \bibnamefont{Graham}},
  \bibinfo{author}{\bibfnamefont{P.~C.} \bibnamefont{Riedi}}, \bibnamefont{and}
  \bibinfo{author}{\bibfnamefont{B.~M.} \bibnamefont{Wanklyn}},
  \bibinfo{journal}{J. Phys. Condens. Matter} \textbf{\bibinfo{volume}{3}},
  \bibinfo{pages}{135} (\bibinfo{year}{1991}).

\bibitem[{\citenamefont{O'Keeffe and Stone}(1962)}]{OS62}
\bibinfo{author}{\bibfnamefont{M.}~\bibnamefont{O'Keeffe}} \bibnamefont{and}
  \bibinfo{author}{\bibfnamefont{F.~S.} \bibnamefont{Stone}},
  \bibinfo{journal}{J. Phys. Chem. Solids.} \textbf{\bibinfo{volume}{23}},
  \bibinfo{pages}{261} (\bibinfo{year}{1962}).

\bibitem[{\citenamefont{Roden et~al.}(1987)\citenamefont{Roden, Braun, and
  Freimuth}}]{RBF87}
\bibinfo{author}{\bibfnamefont{B.}~\bibnamefont{Roden}},
  \bibinfo{author}{\bibfnamefont{E.}~\bibnamefont{Braun}}, \bibnamefont{and}
  \bibinfo{author}{\bibfnamefont{A.}~\bibnamefont{Freimuth}},
  \bibinfo{journal}{Solid State Commun.} \textbf{\bibinfo{volume}{64}},
  \bibinfo{pages}{1051} (\bibinfo{year}{1987}).

\bibitem[{\citenamefont{Kondo et~al.}(1988)\citenamefont{Kondo, Ono, Sugiura,
  Sugiyama, and Date}}]{KOS+88}
\bibinfo{author}{\bibfnamefont{O.}~\bibnamefont{Kondo}},
  \bibinfo{author}{\bibfnamefont{M.}~\bibnamefont{Ono}},
  \bibinfo{author}{\bibfnamefont{E.}~\bibnamefont{Sugiura}},
  \bibinfo{author}{\bibfnamefont{K.}~\bibnamefont{Sugiyama}}, \bibnamefont{and}
  \bibinfo{author}{\bibfnamefont{M.}~\bibnamefont{Date}}, \bibinfo{journal}{J.
  Phys. Soc. Jpn.} \textbf{\bibinfo{volume}{57}}, \bibinfo{pages}{3293}
  (\bibinfo{year}{1988}).

\bibitem[{\citenamefont{K{\"{o}}bler and Chattopadhyay}(1991)}]{KC91}
\bibinfo{author}{\bibfnamefont{U.}~\bibnamefont{K{\"{o}}bler}}
  \bibnamefont{and}
  \bibinfo{author}{\bibfnamefont{T.}~\bibnamefont{Chattopadhyay}},
  \bibinfo{journal}{Z. Phys. B} \textbf{\bibinfo{volume}{82}},
  \bibinfo{pages}{383} (\bibinfo{year}{1991}).

\bibitem[{\citenamefont{{Chandrasekhar Rao} and Sahni}(1994)}]{CRS94}
\bibinfo{author}{\bibfnamefont{T.~V.} \bibnamefont{{Chandrasekhar Rao}}}
  \bibnamefont{and} \bibinfo{author}{\bibfnamefont{V.~C.} \bibnamefont{Sahni}},
  \bibinfo{journal}{J. Phys. Condens. Matter} \textbf{\bibinfo{volume}{6}},
  \bibinfo{pages}{L423} (\bibinfo{year}{1994}).

\bibitem[{\citenamefont{Eskes et~al.}(1990)\citenamefont{Eskes, Tjeng, and
  Sawatzky}}]{ETS90}
\bibinfo{author}{\bibfnamefont{E.}~\bibnamefont{Eskes}},
  \bibinfo{author}{\bibfnamefont{L.~H.} \bibnamefont{Tjeng}}, \bibnamefont{and}
  \bibinfo{author}{\bibfnamefont{G.~A.} \bibnamefont{Sawatzky}},
  \bibinfo{journal}{Phys. Rev. B} \textbf{\bibinfo{volume}{41}},
  \bibinfo{pages}{288} (\bibinfo{year}{1990}).

\bibitem[{\citenamefont{Anderson}(1964)}]{And64}
\bibinfo{author}{\bibfnamefont{P.~W.} \bibnamefont{Anderson}}, in
  \emph{\bibinfo{booktitle}{Magnetism}}, edited by
  \bibinfo{editor}{\bibfnamefont{G.}~\bibnamefont{Rado}} \bibnamefont{and}
  \bibinfo{editor}{\bibfnamefont{H.}~\bibnamefont{Suhl}}
  (\bibinfo{publisher}{Academic Press}, \bibinfo{year}{1964}),
  vol.~\bibinfo{volume}{1}, chap.~\bibinfo{chapter}{2}.

\bibitem[{\citenamefont{Hay et~al.}(1975)\citenamefont{Hay, Thibeault, and
  Hoffmann}}]{HTH75}
\bibinfo{author}{\bibfnamefont{J.}~\bibnamefont{Hay}},
  \bibinfo{author}{\bibfnamefont{J.}~\bibnamefont{Thibeault}},
  \bibnamefont{and} \bibinfo{author}{\bibfnamefont{R.}~\bibnamefont{Hoffmann}},
  \bibinfo{journal}{J. Am. Chem. Soc.} \textbf{\bibinfo{volume}{97}},
  \bibinfo{pages}{4884} (\bibinfo{year}{1975}).

\bibitem[{\citenamefont{{\AA}sbrink and Waslowska}(1991)}]{AW91}
\bibinfo{author}{\bibfnamefont{S.}~\bibnamefont{{\AA}sbrink}} \bibnamefont{and}
  \bibinfo{author}{\bibfnamefont{A.}~\bibnamefont{Waslowska}},
  \bibinfo{journal}{J. Phys. Condens. Matter} \textbf{\bibinfo{volume}{3}},
  \bibinfo{pages}{8173} (\bibinfo{year}{1991}).

\bibitem[{\citenamefont{{Chandrasekhar Rao}
  et~al.}()\citenamefont{{Chandrasekhar Rao}, Kitazawa, Matsumoto, Naka, and
  Shimizu}}]{CKM+97}
\bibinfo{author}{\bibfnamefont{T.~V.} \bibnamefont{{Chandrasekhar Rao}}},
  \bibinfo{author}{\bibfnamefont{H.}~\bibnamefont{Kitazawa}},
  \bibinfo{author}{\bibfnamefont{T.}~\bibnamefont{Matsumoto}},
  \bibinfo{author}{\bibfnamefont{T.}~\bibnamefont{Naka}}, \bibnamefont{and}
  \bibinfo{author}{\bibfnamefont{T.}~\bibnamefont{Shimizu}},
  \bibinfo{note}{(unpublished)}.

\bibitem[{\citenamefont{Carter et~al.}(1977)\citenamefont{Carter, Bennett, and
  Kahan}}]{CBK}
\bibinfo{author}{\bibfnamefont{G.~C.} \bibnamefont{Carter}},
  \bibinfo{author}{\bibfnamefont{L.~H.} \bibnamefont{Bennett}},
  \bibnamefont{and} \bibinfo{author}{\bibfnamefont{D.~J.} \bibnamefont{Kahan}},
  \emph{\bibinfo{title}{Metallic Shifts in NMR}}, \bibinfo{number}{Part I}
  (\bibinfo{publisher}{Pergamon Press}, \bibinfo{year}{1977}).

\bibitem[{\citenamefont{Clogston et~al.}(1964)\citenamefont{Clogston,
  Jaccarino, and Yafet}}]{CJY64}
\bibinfo{author}{\bibfnamefont{A.~M.} \bibnamefont{Clogston}},
  \bibinfo{author}{\bibfnamefont{V.}~\bibnamefont{Jaccarino}},
  \bibnamefont{and} \bibinfo{author}{\bibfnamefont{Y.}~\bibnamefont{Yafet}},
  \bibinfo{journal}{Phys. Rev.} \textbf{\bibinfo{volume}{134}},
  \bibinfo{pages}{A650} (\bibinfo{year}{1964}).

\bibitem[{\citenamefont{Abragam and Bleaney}(1970)}]{AB}
\bibinfo{author}{\bibfnamefont{A.}~\bibnamefont{Abragam}} \bibnamefont{and}
  \bibinfo{author}{\bibfnamefont{B.}~\bibnamefont{Bleaney}},
  \emph{\bibinfo{title}{Electron Paramagnetic Resonance of Transition Ions}}
  (\bibinfo{publisher}{Oxfor University Press}, \bibinfo{year}{1970}).

\bibitem[{\citenamefont{Hellwege and Hellwege}(1966)}]{LB}
\bibinfo{editor}{\bibfnamefont{K.~H.} \bibnamefont{Hellwege}} \bibnamefont{and}
  \bibinfo{editor}{\bibfnamefont{A.~M.} \bibnamefont{Hellwege}}, eds.,
  \emph{\bibinfo{title}{Landolt-B{\"{o}}rnstein}}, vol.~\bibinfo{volume}{16} of
  \emph{\bibinfo{series}{New Series II}} (\bibinfo{publisher}{Springer-Verlag},
  \bibinfo{year}{1966}).

\bibitem[{\citenamefont{Shimizu}(1993)}]{Shi93}
\bibinfo{author}{\bibfnamefont{T.}~\bibnamefont{Shimizu}}, \bibinfo{journal}{J.
  Phys. Soc. Jpn.} \textbf{\bibinfo{volume}{62}}, \bibinfo{pages}{772}
  (\bibinfo{year}{1993}).

\bibitem[{\citenamefont{Shimizu et~al.}(1993)\citenamefont{Shimizu, Aoki,
  Yasuoka, Tsuda, Ueda, Yoshimura, and Kosuge}}]{SAY+93}
\bibinfo{author}{\bibfnamefont{T.}~\bibnamefont{Shimizu}},
  \bibinfo{author}{\bibfnamefont{H.}~\bibnamefont{Aoki}},
  \bibinfo{author}{\bibfnamefont{H.}~\bibnamefont{Yasuoka}},
  \bibinfo{author}{\bibfnamefont{T.}~\bibnamefont{Tsuda}},
  \bibinfo{author}{\bibfnamefont{Y.}~\bibnamefont{Ueda}},
  \bibinfo{author}{\bibfnamefont{K.}~\bibnamefont{Yoshimura}},
  \bibnamefont{and} \bibinfo{author}{\bibfnamefont{K.}~\bibnamefont{Kosuge}},
  \bibinfo{journal}{J. Phys. Soc. Jpn.} \textbf{\bibinfo{volume}{62}},
  \bibinfo{pages}{3710} (\bibinfo{year}{1993}), \bibinfo{note}{and references
  therein.}

\bibitem[{\citenamefont{Eggert et~al.}(1994)\citenamefont{Eggert, Affleck, and
  Takahashi}}]{EAT94}
\bibinfo{author}{\bibfnamefont{S.}~\bibnamefont{Eggert}},
  \bibinfo{author}{\bibfnamefont{I.}~\bibnamefont{Affleck}}, \bibnamefont{and}
  \bibinfo{author}{\bibfnamefont{M.}~\bibnamefont{Takahashi}},
  \bibinfo{journal}{Phys. Rev. Lett.} \textbf{\bibinfo{volume}{73}},
  \bibinfo{pages}{332} (\bibinfo{year}{1994}).

\bibitem[{\citenamefont{Okabe et~al.}(1988)\citenamefont{Okabe, Kikuchi, and
  Nagi}}]{OKN88}
\bibinfo{author}{\bibfnamefont{Y.}~\bibnamefont{Okabe}},
  \bibinfo{author}{\bibfnamefont{M.}~\bibnamefont{Kikuchi}}, \bibnamefont{and}
  \bibinfo{author}{\bibfnamefont{A.~D.~S.} \bibnamefont{Nagi}},
  \bibinfo{journal}{Phys. Rev. Lett.} \textbf{\bibinfo{volume}{61}},
  \bibinfo{pages}{2971} (\bibinfo{year}{1988}).

\bibitem[{\citenamefont{Hidaka et~al.}(1997)\citenamefont{Hidaka, Hatae,
  Yamada, Nishi, and Akimitsu}}]{HHY+97}
\bibinfo{author}{\bibfnamefont{M.}~\bibnamefont{Hidaka}},
  \bibinfo{author}{\bibfnamefont{M.}~\bibnamefont{Hatae}},
  \bibinfo{author}{\bibfnamefont{I.}~\bibnamefont{Yamada}},
  \bibinfo{author}{\bibfnamefont{M.}~\bibnamefont{Nishi}}, \bibnamefont{and}
  \bibinfo{author}{\bibfnamefont{J.}~\bibnamefont{Akimitsu}},
  \bibinfo{journal}{J. Phys.: Condens. Matter} \textbf{\bibinfo{volume}{9}},
  \bibinfo{pages}{809} (\bibinfo{year}{1997}).

\bibitem[{\citenamefont{Pickett}(1989)}]{Pic89}
\bibinfo{author}{\bibfnamefont{W.~E.} \bibnamefont{Pickett}},
  \bibinfo{journal}{Rev. Mod. Phys.} \textbf{\bibinfo{volume}{61}},
  \bibinfo{pages}{433} (\bibinfo{year}{1989}), \bibinfo{note}{and references
  therein}.

\bibitem[{\citenamefont{Fabricius et~al.}(1998)\citenamefont{Fabricius,
  Klumper, Low, Buchner, Lorenz, Dhalenne, and Revcolevschi}}]{FKL+98}
\bibinfo{author}{\bibfnamefont{K.}~\bibnamefont{Fabricius}},
  \bibinfo{author}{\bibfnamefont{A.}~\bibnamefont{Klumper}},
  \bibinfo{author}{\bibfnamefont{U.}~\bibnamefont{Low}},
  \bibinfo{author}{\bibfnamefont{B.}~\bibnamefont{Buchner}},
  \bibinfo{author}{\bibfnamefont{T.}~\bibnamefont{Lorenz}},
  \bibinfo{author}{\bibfnamefont{G.}~\bibnamefont{Dhalenne}}, \bibnamefont{and}
  \bibinfo{author}{\bibfnamefont{A.}~\bibnamefont{Revcolevschi}},
  \bibinfo{journal}{Phys. Rev. B} \textbf{\bibinfo{volume}{57}},
  \bibinfo{pages}{1102} (\bibinfo{year}{1998}).

\bibitem[{\citenamefont{Nishi et~al.}(1994)\citenamefont{Nishi, Fujita, and
  Akimitsu}}]{NFA94}
\bibinfo{author}{\bibfnamefont{M.}~\bibnamefont{Nishi}},
  \bibinfo{author}{\bibfnamefont{O.}~\bibnamefont{Fujita}}, \bibnamefont{and}
  \bibinfo{author}{\bibfnamefont{J.}~\bibnamefont{Akimitsu}},
  \bibinfo{journal}{Phys. Rev. B} \textbf{\bibinfo{volume}{50}},
  \bibinfo{pages}{6508} (\bibinfo{year}{1994}).

\bibitem[{\citenamefont{Shamoto et~al.}(1993)\citenamefont{Shamoto, Sato,
  Tranquada, Sternlieb, and Shirane}}]{SST+93}
\bibinfo{author}{\bibfnamefont{S.}~\bibnamefont{Shamoto}},
  \bibinfo{author}{\bibfnamefont{M.}~\bibnamefont{Sato}},
  \bibinfo{author}{\bibfnamefont{J.~M.} \bibnamefont{Tranquada}},
  \bibinfo{author}{\bibfnamefont{B.~J.} \bibnamefont{Sternlieb}},
  \bibnamefont{and} \bibinfo{author}{\bibfnamefont{G.}~\bibnamefont{Shirane}},
  \bibinfo{journal}{Phys. Rev. B} \textbf{\bibinfo{volume}{48}},
  \bibinfo{pages}{13817} (\bibinfo{year}{1993}).

\bibitem[{\citenamefont{Lyons et~al.}(1988{\natexlab{a}})\citenamefont{Lyons,
  Fleury, Schneemeyer, and Waszczak}}]{LFSW88}
\bibinfo{author}{\bibfnamefont{K.~B.} \bibnamefont{Lyons}},
  \bibinfo{author}{\bibfnamefont{P.~A.} \bibnamefont{Fleury}},
  \bibinfo{author}{\bibfnamefont{L.~F.} \bibnamefont{Schneemeyer}},
  \bibnamefont{and} \bibinfo{author}{\bibfnamefont{V.}~\bibnamefont{Waszczak}},
  \bibinfo{journal}{Phys. Rev. Lett.} \textbf{\bibinfo{volume}{60}},
  \bibinfo{pages}{732} (\bibinfo{year}{1988}{\natexlab{a}}).

\bibitem[{\citenamefont{Johnston}(1989)}]{Joh89}
\bibinfo{author}{\bibfnamefont{D.~C.} \bibnamefont{Johnston}},
  \bibinfo{journal}{Phys. Rev. Lett.} \textbf{\bibinfo{volume}{62}},
  \bibinfo{pages}{957} (\bibinfo{year}{1989}).

\bibitem[{\citenamefont{Bourges et~al.}(1997)\citenamefont{Bourges, Casalta,
  Ivanov, and Petitgrand}}]{BCIP98}
\bibinfo{author}{\bibfnamefont{P.}~\bibnamefont{Bourges}},
  \bibinfo{author}{\bibfnamefont{H.}~\bibnamefont{Casalta}},
  \bibinfo{author}{\bibfnamefont{A.~S.} \bibnamefont{Ivanov}},
  \bibnamefont{and}
  \bibinfo{author}{\bibfnamefont{D.}~\bibnamefont{Petitgrand}},
  \bibinfo{journal}{Phys. Rev. Lett.} \textbf{\bibinfo{volume}{79}},
  \bibinfo{pages}{4906} (\bibinfo{year}{1997}).

\bibitem[{\citenamefont{Lyons et~al.}(1988{\natexlab{b}})\citenamefont{Lyons,
  Fleury, Rameika, Cooper, and Negran}}]{LFR+88}
\bibinfo{author}{\bibfnamefont{K.~B.} \bibnamefont{Lyons}},
  \bibinfo{author}{\bibfnamefont{P.~A.} \bibnamefont{Fleury}},
  \bibinfo{author}{\bibfnamefont{J.~P.} \bibnamefont{Rameika}},
  \bibinfo{author}{\bibfnamefont{A.~S.} \bibnamefont{Cooper}},
  \bibnamefont{and} \bibinfo{author}{\bibfnamefont{T.~J.}
  \bibnamefont{Negran}}, \bibinfo{journal}{Phys. Rev. B}
  \textbf{\bibinfo{volume}{37}}, \bibinfo{pages}{2353}
  (\bibinfo{year}{1988}{\natexlab{b}}).

\bibitem[{\citenamefont{Mizuno et~al.}(1998)\citenamefont{Mizuno, Tohyama,
  Maekawa, Osafune, Motoyama, Eisaki, and Uchida}}]{MTM+98}
\bibinfo{author}{\bibfnamefont{Y.}~\bibnamefont{Mizuno}},
  \bibinfo{author}{\bibfnamefont{T.}~\bibnamefont{Tohyama}},
  \bibinfo{author}{\bibfnamefont{S.}~\bibnamefont{Maekawa}},
  \bibinfo{author}{\bibfnamefont{T.}~\bibnamefont{Osafune}},
  \bibinfo{author}{\bibfnamefont{N.}~\bibnamefont{Motoyama}},
  \bibinfo{author}{\bibfnamefont{H.}~\bibnamefont{Eisaki}}, \bibnamefont{and}
  \bibinfo{author}{\bibfnamefont{S.}~\bibnamefont{Uchida}},
  \bibinfo{journal}{Phys. Rev. B} \textbf{\bibinfo{volume}{57}},
  \bibinfo{pages}{5326} (\bibinfo{year}{1998}).

\bibitem[{\citenamefont{Yokoyama and Saiga}(1997)}]{YS97}
\bibinfo{author}{\bibfnamefont{H.}~\bibnamefont{Yokoyama}} \bibnamefont{and}
  \bibinfo{author}{\bibfnamefont{Y.}~\bibnamefont{Saiga}}, \bibinfo{journal}{J.
  Phys. Soc. Jpn.} \textbf{\bibinfo{volume}{66}}, \bibinfo{pages}{3617}
  (\bibinfo{year}{1997}).

\end{thebibliography}
\end{document}